\documentclass[onecolumn,aps,prl]{revtex4}
\usepackage{graphicx}
\usepackage{epstopdf}
\usepackage{amsfonts}
[1999/03/29 PSNFSS v.7.2
Times font as default roman
: S Rahtz]

\begin{document}
\renewcommand{\thefootnote}{\fnsymbol{footnote}}
\sloppy
\newcommand{\rp}{\right)}
\newcommand{\lp}{\left(}
\newcommand \be  {\begin{equation}}
\newcommand \ba {\begin{eqnarray}}
\newcommand \bas {\begin{eqnarray*}}
\newcommand \ee  {\end{equation}}
\newcommand \ea {\end{eqnarray}}
\newcommand \eas {\end{eqnarray*}}

\title{A theoretical framework for trading experiments}
\thispagestyle{empty}

\author{Maxence Soumare$^1$, J\o rgen Vitting Andersen$^{2,4}$, 
Francis Bouchard$^3$,  
Alain Elkaim$^3$,  
Dominique Gu\'egan$^2$,  
Justin Leroux$^3$,  
Michel Miniconi$^1$, 
and Lars Stentoft$^3$  
\vspace{0.5cm}}
\affiliation{$^1$Laboratoire J.-A. Dieudonn\'e Universit\'e de Nice-Sophia 
Antipolis, Parc Valrose 06108 Nice Cedex 02, France\\}
\affiliation{$^2$CNRS, Centre d'Economie de la Sorbonne, Universit\'e Paris 1 Panth\'eon-Sorbonne, 
Maison des Sciences Economiques, 106-112 Boulevard de l'H\^opital 75647 Paris Cedex 13
, France\\} 
\affiliation{$^3$HEC Montr\'eal, 3000 chemin de la C\^ote Sainte-Catherine, Montr\'eal, 
QC H3T 2A7, Canada
\\}
\affiliation{$^4$Corresponding author; email: jorgen.vitting.andersen@gmail.com, tel: +33 (0)
1 44 07 81 89; fax: +33 (0)1 44 07 81 18\\}
\email{Jorgen-Vitting.Andersen@univ-paris1.fr}


\date{\today}

\begin{abstract}
{\bf 
A general framework is suggested to describe human decision making
in a certain class of experiments performed in a trading laboratory. We are in
particular interested in discerning between two different moods, or states
of the investors, corresponding to investors using fundamental investment
strategies, technical analysis investment strategies 
respectively. Our framework 
accounts for two opposite situations 
already encountered in experimental setups: i) the rational expectations case,
and ii) the case of pure speculation. We consider new experimental
conditions which allow both elements to be present in the decision
making process of the traders, thereby creating a dilemma in terms of investment strategy. Our 
theoretical framework allows us to predict the outcome of this 
type of trading experiments, depending 
on such variables as the number of people trading, the liquidity of the 
market, the amount of information used in technical analysis strategies, as well as 
the dividends attributed to an asset. We find that it is possible to give 
a qualitative prediction of trading behavior depending on a ratio that 
quantifies the fluctuations in the model.  \\ 

}
\end{abstract}

\maketitle

\vspace{1cm}

\section{1. Introduction}

In order to gain new insight on how investors perceive investment
possibilities as well as risks in financial markets, it appears important to
confirm not only the background of theoretical studies on human decision
making, but also to get knowledge from controlled experiments, where one can
probe in detail the different assumptions of investment behavior.
It is only recently that experimental Finance has begun to appear as a
well-established field, the interest in particular sparked by the recognition
in terms of the attribution of the Nobel Prize in Economics to Vernon Smith in
2002. However, so far the major part of experimental work in Finance has
assumed (Vernon Smith included) human rationality and the ability of
markets to find the proper price close to an equilibrium setting. Contrary to
this approach Behavioural Finance takes a more practitioner-minded
description of how actual decision making takes place in financial markets. It
would therefore seem like a very natural approach to bridge the insight gained
from Behavioral Finance and apply it to experiments done on financial
markets. Interestingly, not much effort has been done in this direction. The
main reason is maybe because the major part of research done in Behavioural
Finance is concerned with how \emph{individual} decision making takes place
(Prospect Theory included (Tversky and Kahneman (1974), Kahneman and Tversky (1979), 
and Tversky and Kahneman (1991) )) and in a \emph{static} setting, whereas price
setting in financial markets is clearly an \emph{aggregate}
and \emph{dynamic} phenomenon.

The efforts in this paper should be seen from such a viewpoint. Our
theoretical foundation is based on a complex systems approach that places
emphasis on social learning and group behaviour in order to understand the
price formation in financial markets. The idea is that financial market
participants are connected through their impact on the price as well as
through the percolation of information through the group of market
participants. 
For instance,  \textquotedblleft
shocks\textquotedblright\ created by a large liquidation of a given market
participant can have future impacts on the decision making of other market
participants who in turn follow a similar decision to liquidate their
positions. In the context where both dynamic behaviour as well as social
learning and group behaviour are relevant, tools from
complexity theory are particularly appealing, including for 
example agent-based modeling and game theory as presented here. 

\bigskip

We are especially interested in discerning
between two different moods (or states) of the population of investors, corresponding to
i) investors using fundamental investment strategies as in the case of 
rational expectations and ii) the emergence of a speculative bias as seen 
in certain cases when investors use technical analysis strategies.
The rational expectations case i) has been studied extensively in a large
number of experiments under various
situations and with different constraints (Smith (1962), Smith (1965), Plott and Smith (1978), Coppinger et al. (1980), Hommes et al. (2008)). In the simplest setup, which is
included in our theoretical description further below, people trade shares of
a given company based on their expectations of future dividends of the
company. Throughout the experiment such expectations change due to the arrival
of new information. The experiment ends with the closure of the company and
the payout of the dividends to the participants in the experiments. It should
be noted that in this case there is no incentive for the participants to
speculate on the price itself since the full price of the company reflects the
expected dividends payout \emph{at the end of the experiment}. A case study
was done for the opposite situation where expectations about dividends do
\emph{not} play any role, and reported in Roszczynska-Kurasinska et al. (2012). In this experiment
only the price was available for the investment decisions of the group of
participants. However 
 as was shown in 
Roszczynska-Kurasinska et al. (2012)
, it requires coordination among the  
participants to profit from a speculation bias 
 in this kind of experiments. 

\bigskip

It should be noted that our approach differs from most of the 
Behavioral Finance/bounded-rationality literature in that
the phenomena we study can only be understood by looking at the system level.
In other words, although the phenomena that emerge depend on microscopic
features of the agents, it is important to not only look at individual
characteristics but to study the system as a whole. The state of the system
(speculative or fundamentalist) is the macroscopic result of many microscopic
decisions. We shall refer to this collective \textquotedblleft
choice\textquotedblright\ of the state of the system as \textquotedblleft
aggregate decision making\textquotedblright.

Our setup is conducive to answering a number of well-defined questions--that
of how prices form, for instance-- that the Rational Expectations Hypothesis
(REH), or even the bounded REH cannot address. Once one mixes in the two
ingredients that are the asset price and the dividends, the situation becomes
ambiguous: REH suggests agents should make their trading decisions based on
dividends, but the matter becomes far from trivial once the price of an asset
and an end date are factored in. Are decisions based on the price of the asset
in anticipation of future price behavior (speculative state), or are dividends
the only drivers of agents' trading decisions (fundamentalist state)?\bigskip

In the following we introduce theoretical foundations 
encompassing the occurrence of both the
speculative and the fundamentalist state. We do so by considering the "Dollar
Game" (or "\$-Game"), which is an investment game that combines the two key
ingredients that are dividends and the asset price (Vitting Andersen, Sornette (2003))
. Although simple in
principle, the \$-Game yields rich system dynamics, the complexity of which
can be acted upon by the choice of system parameters (memory length,
liquidity, etc.). As will be seen, this thereby creates a dilemma in terms of
the investment strategies of the participants. The pure cases i) and ii) will
appear as special cases of the general theory.

This feature of the \$-Game lends itself well to study using a general
well-known theory of phase-transitions found in Physics: the
Ginzburg-Landau theory (henceforth referred to as "the GL theory") which 
we describe later.

\noindent

\section{2. The \$-Game}

The \$-Game was inspired by the Minority Game (MG) introduced in 1997 by Ye-Cheng
Zhang and Damien Challet (Challet and Zhang (1997), Challet and Zhang (1998) as an agent-based model proposed to
study market price dynamics (Zhang (1998), Johnson et al. (1999), Lamper et al. (2002) 
). The MG was introduced
following a leading principle in Physics, that in order to solve a complex
problem one should first identify essential factors at the expense of trying
to describe all aspects in detail. Similar to the Minority Game, the \$-Game
should be considered as a \textquotedblleft minimal\textquotedblright\ model
of a financial market.

Formally, $N$ players (or agents) simultaneously take part in a one-asset
financial market over a horizon of $T$ periods. At each period, $t\leq T$,
each player $i$ chooses an action $a_{i}(t)\in\{-1,1\}$, where action "$1$" is
interpreted as "buy" and action "-1" as "sell". 
Players are assumed to be boundedly rational, in the sense
of using only a limited information set upon which to base decisions. In the version of 
the \$-Game presented in this paper, the agents use two different types of 
investment strategies, technical analysis strategies and fundamental analysis 
strategies. 
Concerning  
the decision making related to technical analysis, each
player observes the history of past price movements, which is limited to the size of
their memory, 
$m \in\mathbb{N}$.\ Each player has at his/her disposal a fixed number of $s$  
strategies which 
are randomly assigned at the beginning of the game.   
It follows that player $i$'s, $j$'th 
strategy, $a_{i}^j$, is a mapping from the set of histories of size
$m$ to $\{-1,1\}$. We denote by $\vec{h}(t)\in\{0,1\}^{m}$ the history
vector that agents observe in period $t$ before taking the action of either buying or 
selling an asset. We
interpret "1" to represent an up move of the market (an increase in the asset
price) and "0" corresponds to a down move of the market (a fall in the price
of the asset). These assumptions are equivalent to having agents behave as
technical analysts who use lookup tables to determine their next move. Table~1
 shows an example of a strategy for $m=3$:%

\begin{table}
\begin{center}
\begin{tabular}
[c]{|c|c|}\hline\hline
history, $\vec{h}(t)$ & action, $a_{i}^j(t)$\\\hline
000 & 1\\
001 & -1\\
010 & -1\\
011 & 1\\
100 & -1\\
101 & -1\\
110 & 1\\
111 & -1\\\hline\hline
\end{tabular}
\end{center}
\label{table_MG}
\text{Example of a strategy}\nonumber
\end{table}
A strategy therefore tells an agent what to do given the past market behavior. If
the market went down over the last three days, the strategy represented in
Table~1 suggests that now is a good moment to buy
($000\rightarrow1$) in Table~1. If instead the market went down
over the last two days and then up today, the same strategy suggests that now
is a good moment to sell ($001\rightarrow-1$) in Table~1. 
While a single strategy recommends an action for all possible histories (of length $m$), 
 we also allow for agents to adopt different strategies over time. Namely, agents keep a 
record of the overall payoff each strategy would have yielded over the entire market history 
(i.e. not limited to $m$ periods prior) and use this record to update which strategy is 
the most profitable. 
In every 
time period agent $i$ therefore choses 
the {\bf best} strategy (in terms of payoff, see definition 
below)  out of the $s$ available. This 
renders the game highly non-linear:  as the price behavior of the market 
changes, the best strategy of a given agent changes, which then can lead to new 
changes in the price dynamics. The action of the best strategy of agent $i$ at time 
$t$, is denoted by $a^*_i(t)$. 
We
denote by $\left(  a^*(t)\right)  _{i}\in\{-1,1\}^{N\times T}$\ the \emph{action
profile} of the population, where $\vec{a}^*(t)=\left(  a^*_{1}(t),...,a^*_{N}(t)\right)
\in\{-1,1\}^{N}$ corresponds to the action played by the $N$ agents in period
$t$.

The payoff $\pi$ of the $i$th agent's $j$th strategy, $a_i^j$,  
in period $t$ is determined as follows:
\begin{equation}
\pi [ a_i^j ]  =a_{i}^j (t-1)  \sum_{k=1}^{N}a_k^* (
t )  \label{payoff_DG}%
\end{equation}
The return $r(t)$ of the market between period $t$ and $t+1$ is assumed to be proportional to 
the order imbalance 
$\sum_{k=1}^{N}a_k^* (t)$: 
\be  
r(t)  = 
\sum_{k=1}^{N}a_k^* (t)/\lambda  
\label{r_DG}
\ee 
with $\lambda$ a parameter describing the liquidity of the market.  
Therefore the payoff of a given strategy (\ref{payoff_DG}) can be 
expressed in terms of the return of the market in the next time period as: 
\begin{equation}
\pi [ a_{i}^j ]  =a_{i}^j\left(  t-1\right) \lambda r(t)
\label{payoff_DG_1}
\end{equation}

From (\ref{payoff_DG}) one can see that the payoff depends on two different times: the individual
decision at time $t-1$ and the aggregate "decision" at time $t$. Such a feature
of the payoff function is illustrative of real financial markets, where
traders decide to enter a position in a market at time $t-1$, 
but do not know their return until the
market closes the next day (time $t$). This is especially clear from
 (\ref{payoff_DG_1}) where it can be seen that the \$-Game rewards a given strategy 
that at time $t-1$ predicted the proper direction of the return of the market $r(t)$ 
in the {\em next} time step 
 $t$. The larger the move of the market, the larger the gain/loss depending on whether 
the strategy properly/improperly predicted the market move.  
Therefore in the \$-Game, agents correspond to
speculators trying to profit from predicting the direction of price change.

In addition to technical analysis strategies that try to profit from price changes, we also consider 
strategies that try to profit from information of the fundamental value of an asset $P_f(t)$.  
$P_f(t)$ is determined entirely from future expectations about the dividends $d(t)$ attributed to the 
asset at the end of the experiment. Whenever $P(t) > P_f(t)$  a fundamental strategy therefore 
gives the recommendation to sell, whereas if  
$P(t) < P_f(t)$  it recommends buying. Furthermore in order to take into account a diminishing 
use of such strategies in a purely speculative phase when the price $P(t) >> P_f(t)$, the 
probability to use a fundamental strategy is taken from a Poisson distribution 
$\gamma \exp{(-\gamma )}$ with $\gamma = {P(t)-P_f \over d}$.

To sum up, the \$-Game as described in this article can be described in 
terms of just five parameters:

\begin{itemize}
\item $N$ - The number of agents (market participants).

\item $m$ - The memory length used by the agents.

\item $s$ - The number of strategies held by the agents. It should be noted that the $s$ 
strategies of each agent is chosen randomly (corresponding to a random column of `$0$'s 
and $1$'s in table~1) in the total pool of $2^{2^m}$ strategies at the 
beginning of the game.  

\item $\lambda$ - The liquidity parameter of the market.

\item $d(t)$ - The future expectations about the dividends paid at 
the end of the experiment. To simplify,  $d(t)$ will be taken constant in time $t$ 
in this paper.  
\end{itemize}

The dynamics of the \$-Game are driven by nonlinear feedback
 because each agent uses his/her
\emph{best} strategy at every time step. 
 As the market changes,
the best strategies of the agents change, and as the strategies of the agents
change, they thereby change the market. Formally one can understand such  
dynamics by 
representing the price history $h(t)  =   \sum_{j=1}^{m} b(t-j+1) 2^{j-1}$ as a 
scalar where $b(t)$ is the bit representing the direction of price movement 
at time $t$ (see table~1). The dynamics of the \$-Game can then be 
expressed in terms of an equation that describes the dynamics of $b(t)$ as: 
\ba
b(t+1)  =   \Theta (\sum_{i=1}^{N} a^*_i (h (t)) ), & & 
\label{def_b}
\ea
with $\Theta$ a Heaviside function. 
The nonlinearity of the game can be formally seen from: 
\ba
 a^*_i (h (t) )  =  
a_i^{\{j |  {\rm max}_{j=1,...,s} \{ \Pi [a_i^j (h (t) )  ]\} } (h(t)),  & & 
\Pi  [a_i^j (h (t) ) ]  =  \sum_{k=1}^t  a_i^j(h (k-1)) \sum_{i=1}^N a^*_i(h(k)) 
\label{def_a_star}
\ea

Inserting the expressions (\ref{def_a_star}) in expression  
 (\ref{def_b}) one obtains an expression that describes the \$-Game  
 in terms of just one single equation for $b(t)$ depending on the 
values of the variables $(m,s,N,\lambda,d)$ and the random variables $a_i^j$ 
(i.e. their initial random assignments).  
A major complication in the study of this equation happens because 
of the non-linearity in the selection of the best strategy. For 
$s=2$ however the expressions simplifies because one only need to 
know the relative payoff $q_i \equiv \pi [a_i^1] - \pi[a_i^2]$ between two strategies 
(Challet, Marsili (1999), Challet, Marsili (2001) ). 
For this special case it was shown in 
Roszczynska-Kurasinska et al. (2012)
 that the Nash
equilibrium for the \$-Game with only technical analysis 
strategies (with no cash nor asset constraints) is akin to that
of Keynes' \textquotedblleft Beauty Contest\textquotedblright\ where it
becomes profitable for the subjects to guess the actions of the other
participants. The optimal state is then one for which all subjects cooperate
and take the \emph{same} decision (either buy/sell). 

\section{3. Ginzburg-Landau theory}

To describe further the competition between technical analysis 
trading strategies and fundamental analysis trading strategies as used  in 
the \$-Game, we suggest to borrow a description from Physics where different states of 
a system can be characterized via a so-called free energy $F$. 
$F$ in that case plays a central role, since its minimum
determines how the state of the system will appear. $F$ can be written as $F = E - T
S$ with $E$ the energy of the system, $T$ the temperature and $S$ the entropy
which one can think of as representing how much disorder there is in a given
system. From the definition of $F$ we can see that the state of a system is
determined by a struggle between two different forces, one representing
``order'', this is the $E$ term, and the other term representing ``disorder''
given by the $TS$ term. We suggest a similar struggle of
``forces'' to be present in the trading experiments.

The competition between order and disorder as described by $F$, can be understood in more detail by
considering the example given by the Ising model, which is a model of
ferromagnetism. For the Ising model the energy $E = -J \sum_{<i,j>}  s_{i}
s_{j}$ with $s_{i}, s_{j}$ representing the atomic ``spins'' of a material.
The $<>$-notation in the  summation indicates that the sum
 is to be taken over all nearest neighbors pair of spins. Each
spin itself can be thought of as a mini magnet. In the two-dimensional Ising
model the spin $s_{i}=1$ if the spin is ``up'' and $s_{i}=-1$ if the spin is
``down''. Taking the coupling strength between spins $J$ positive, the minimum
energy $E_{\mathrm{min}}$ of the system is simply given by either all spins up
$(s_{i} \equiv1)$, or all spins down $(s_{i} \equiv-1)$. For temperature $T=0$
the minimum of the energy $E$ is therefore also the minimum of the free energy
$F$. However as soon as $T>0$, the finite temperature will introduce
fluctuations of the spins introducing thereby a non-zero contribution to the
entropy $S$. The larger the temperature $T$ the larger this tendency, until at
a certain temperature $T_{c}$ above which order has completely disappeared -
the system is in a disordered state. Order in the case of the Ising model is
measured by the magnetism, which is just the averaged value of the spin $m =
E(s_{i})$.

GL theory introduces the idea that we can in general understand
such order-disorder transitions mentioned above by expanding the free energy
in terms of the order parameter $m$. Specifically write:
\begin{equation}
F=C+am+\alpha m^{2}+bm^{3}+\beta/2m^{4}+...\label{F_GL}%
\end{equation}
Using now the symmetry argument that there should be no difference in the free
energy between the two states with respectively either all $s_{i}\equiv1$ or
$s_{i}\equiv-1$, all odd order terms in $m$ disappear in (\ref{F_GL}). Taking
furthermore the derivative (in order to find its extreme) we end up with the
equation at a minimum of $F$:
\begin{equation}
0=m(\alpha+\beta m^{2})\label{m_sol}%
\end{equation}
(\ref{m_sol}) has the trivial solution of the magnetization $m=0$, this is the
high temperature solution and describe the disordered state. Taking $\beta$
positive, the other non-trivial solution happens for negative $\alpha$,
$m^{2}=-\alpha/\beta$. By writing $\alpha=(T-T_{C})$ one sees that the
magnetization scales as $(T-Tc)^{1/2}$ for temperatures below $T_{c} $. The
exponent of 0.5 is the so called \textquotedblleft mean
field\textquotedblright\ or GL exponent of the transition.

We now propose to consider in similar terms the competition in the trading
experiments between profit from speculation obtained through trend-following,
versus the tendency to destroy such trends due to mean reversion towards 
the fundamental price (Vitting Andersen (2010). The general tendency to create either a
positive/negative price trend corresponds to \textquotedblleft
order\textquotedblright\ whereas either the lack of consensus or the mean
reversion to the fundamental price value will destroy such order. To make the
analogy with our discussion above, we introduce what one could call the
\textquotedblleft free profit\textquotedblright\ given by two terms
$F_{P}=P-TS$. $P$ is the profit of the ordered state which for $T=0$
corresponds to a continuous up/down trend of the market. $S$ is an entropy term
that destroys the ordered state, and $T$ is the \textquotedblleft
temperature\textquotedblright\,  which will be introduced below.
 
As discussed beforehand, the payoff of a strategy in the \$-Game describes the profit for the given 
strategy.  
Agents in a Nash equilibrium are characterized by using the same 
strategy over time, therefore such a strategy has to be optimal. 
We can then write the total profit $P$ for the 
system of traders in a Nash equilibrium of the \$-Game as:  
\ba
P (t)  & =  &
 \sum_i \pi [a_i^*]  \\
& &   = \sum_{i=1}^N \sum_{j=1}^{N} a_i^*(t-1) a_j^* (t)
\label{P_DG}
\ea
We note the resemblance of (\ref{P_DG}) to the Ising model described above. One  major  
difference with respect to the Ising model however is the ``interaction''  between 
traders, since 
 (\ref{P_DG}) says that trader $j$'s action at time $t$ has an impact 
on trader $i$'s profit from the action he/she took at time $t-1$. 
 Therefore the ``interaction'' 
is seen to be ``long-ranged'' in (\ref{P_DG}) whereas the interaction is local (it only concerns 
nearest-neighbors) for the Ising model.

Similar to
(\ref{F_GL}) we can introduce an order parameter and expand the
\textquotedblleft free profit\textquotedblright\ in terms of this parameter.
In the case of the Ising model the order parameter is given by
 the spatial average of the local order (the magnetization). 
In the trading setup we suggest to consider 
the local order $o$ expressed in terms of the order imbalance: $o = {1/N \sum_{i=1}^N a_i^*}$ 
which from (\ref{r_DG}) is seen to be proportional to the return. 
The parameter $o$ varies between -1 (all agents decide to sell) and  1 (all agents decide 
to buy).
In the case where one can neglect the dividends (the
experiment described in 
Roszczynska-Kurasinska et al. (2012)
 $o\rightarrow\pm1$ so this corresponds
to the complete ordered outcome. For the experiments performed under the
assumptions of rational expectations the price converges to the fundamental
price in the end of the experiments and we get $o\rightarrow0$ for
$t\rightarrow t_{n}$ with $t_n$ the duration of the experiment. 

Applying now the GL idea and expanding $F_{p}$ in terms of $o$, 
  one ends up with the very same conditions (\ref{m_sol}) to
determine $o$, except that the extreme (extremes) now describes a maximum
(maxima) instead of a minimum (minima) as was the case for $F$. Note that all
odd order terms of $o$ disappear since there is no difference in the profit
that traders obtain in shorting the market compared to going long. Figure~1
illustrates the expansion of $F_{p}$ ($y$-axis) as a function of $o$
($x$-axis) for the two cases: i) the $T > T_{c}$ solution (i.e. the disordered
state corresponding to no trend in the experiments $o=0$) can be seen as the
maximum of the solid line, whereas the two $T < T_{c}$ solutions (i.e. the
ordered state corresponding to a certain trend in the experiments $o \neq0$)
can be found as the maxima of the dashed line.

One of the main implications of the GL theory is  the 
existence of a nontrivial transition from a high ``temperature'' disordered
state in the trading experiments where traders don't create a trend over time,
to a low `temperature'' state characterized by trend following. 
A ``temperature'' can now be defined via the randomness of the model as will 
be explained in the following. Randomness enters    
 the \$-Game through the initial conditions in the 
assignments of the $s$ strategies to the $N$ traders in the game. In order to create a 
given strategy one has to assign randomly either a $0$ or a $1$ for each of the $2^m$ different 
price histories. Therefore the total pool of 
strategies increases as $2^{2^m}$ versus $m$. However many of these strategies 
are closely related - take e.g. table~1 and change just one of the $0$'s to a $1$, this  
thereby creates a strategy which is highly correlated to the one seen in table~1.   
In Challet, Zhang (1997), Challet, Zhang (1998), 
it was shown how to construct  
 a small subset of size $2^m$ of independent strategies out of the total pool 
of $2^{2^m}$ strategies. As suggested 
in (Savit et al. (1999) and Challet, Marsili (2000)) ,
 a qualitative understanding of the MG  can then be obtained by considering the 
parameter $\alpha  \equiv {2^m \over N }$. However as pointed out in Zhang (1998)   
the  ratio  
 $\alpha^{'} = {2^m \over N \times s}$  seems intuitively to be more relevant since this 
quantity describes  
ratio of the total number of relevant strategies to the total number of strategies held by 
the traders. Taking into account the presence of a fundamental value strategy we therefore 
introduce the ratio $T = {2^m + 1 \over N \times s}$ to describe the temperature in the 
simulations of the \$-Game presented in the following. The relation of $T$ to the fluctuations 
of the system becomes clear when one consider that when sampling the variance of a small  
sample is larger than the variance of a large sample (a fact called ``the law of {\em small} 
numbers'' in Psychology/Behavioral Finance (Tversky, Kahneman (1974), Kahneman, Tversky 
(1979), Tversky, Kahneman (1991) ). Therefore when the sample of strategies 
held by the $N$ traders is small with respect to the total pool of relevant strategies, this 
corresponds to the large fluctuations, large temperature case. Vice versa a large 
sample of strategies held by the $N$ traders therefore corresponds to a small temperature case 
as seen from the definition of $T$.

\section{4. Results}
Figure~2 shows three different results representing typical market behavior corresponding to fundamental price 
behavior, as well as speculative behavior in an increasing/decreasing market.   
Figure~3-4 show histograms representing respectively speculative behavior (blue) or  
fundamentalist  behavior (red) as outcomes in a setup of the \$-Game corresponding to a 
trading experiment with given 
realizations of the 5 parameters $(N,m,s,\lambda,d)$. The histograms in figure~2 
represents simulations performed with $s=2$ whereas the histograms in figure~3 were 
done for simulations with $s=18$.
The different histograms were 
obtained from an ensemble average of 200 simulations of the \$-Game where each realization 
of the game were run for up to $200*{2^m}$ time steps. A speculative state was determined 
whenever $m$ successive price changes had occurred whereas a fundamental state was characterized 
by the price fluctuation within a 50 percentage range of the fundamental value $P_f$. White in 
the figures represents the 
cases where neither a definite speculative nor fundamental state could be defined. 

We first notice the somewhat surprising fact that the dividends $d$ 
as well as the liquidity of the market $\lambda$, only 
seem to have  
 a quite limited impact on 
 the final state of the market. In particular for the smallest $m$ values ($m=3,5$) increasing 
dividends appear to have a somewhat stabilizing effect allowing for slightly more fundamental value 
states. The same stabilizing trend appears to be at play as one 
increases the liquidity of the market, but again, this tendency appear to be very weak. A much 
clearer tendency is seen with    
respect to increasing speculation 
when increasing the number of traders $N$, respectively decreasing the amount of information $m$ 
used in the decision making of the technical analysis trading strategies. A larger number 
of strategies $s$ assigned to the traders is also seen to enhance speculation (compare figure~3 
and figure~4). 

One of our main results is that  
 a qualitative behavior of a trading experiment can  be predicted 
 depending on the 
given value of $T$. In particular with respect to expectations about the outcome in an 
experimental setup of the market model, such an understanding is important (Bouchard et 
al. 2012). 
The fact that $T$ determines the outcome of trading behavior can  be  seen by changing 
the nominator and denominator by the same factor, which  then should lead to invariant behavior 
in terms of trading decisions. This means that for example the 
 $(m=3, N=11)$ (i.e. $T=0.72/s$) trading behavior for a given $\lambda$ and $s$  
should fall in between  the $(m=5,  N=101)$ (i.e. $T=0.32/s$) and $(m=8, N=101)$  (i.e. 
$T=1.27/s$) cases. From figure~2 and figure~3 this is seen indeed to be the case. Similarly 
comparing figure~3-4 it is seen that increasing (/decreasing) $N$ and decreasing (/increasing) 
$s$ by the same amount leads to two systems behaving similarly in terms of investment profile (compare 
$N=101$ rows in figure~3 to $N=11$ rows in figure~4) . 
These results underscore the importance of the  parameter $T$ when it comes to 
the understanding of the aggregate decision making in the model.

\section{5. Conclusion}
A general framework has been  suggested to describe the human decision making
in a certain class of experiments performed in a trading laboratory. 
Our 
framework allows us to predict the outcome of such type of trading experiments in terms 
of when to expect a fundamental versus a speculative state. We have shown how a 
qualitative understanding 
could be found depending  
on just one parameter,  representing the fluctuations of the model. Our findings give 
certain guidance with respect to the implementation of trading experiments performed 
in a trading laboratory (Bouchard et al. (2012)). 

\section{6. Acknowledgments}
J.V.A., M. M.  and M. S. would like to thank the Coll\`ege Interdisciplinaire de la Finance for financial 
support.

{}

\begin{figure}[h]
\includegraphics[width=14cm]{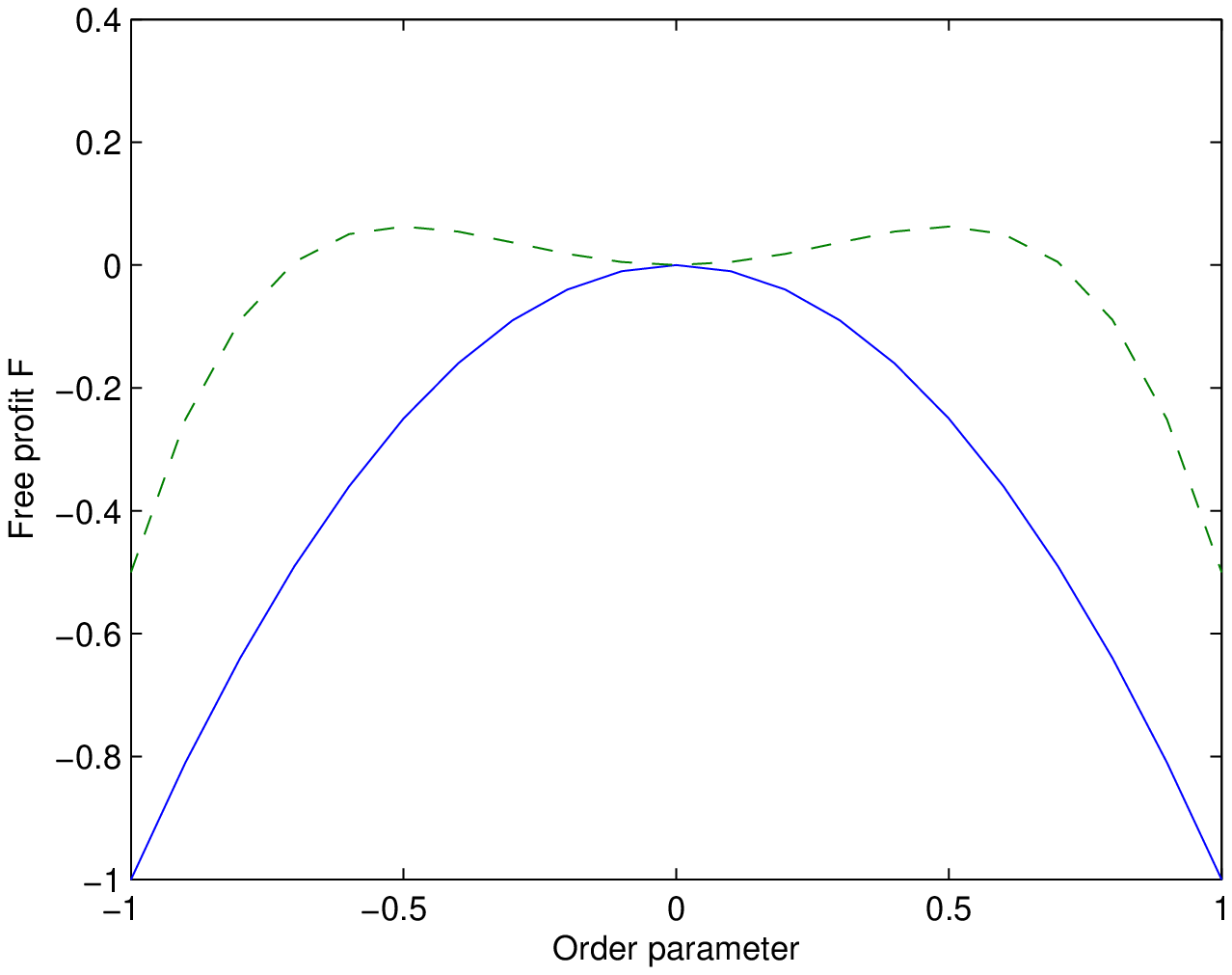}
\caption{\protect\label{Fig1}
Illustration of the ``Free Profit'' $F_p$ as a function of the order parameter $o$ for two 
different ``temperuatures'' corresponding to $
T>T_c$ (solid blue line), and $T<T_c$ (dashed green line) respectively. 
}
\end{figure}

\begin{figure}[h]
\includegraphics[width=14cm]{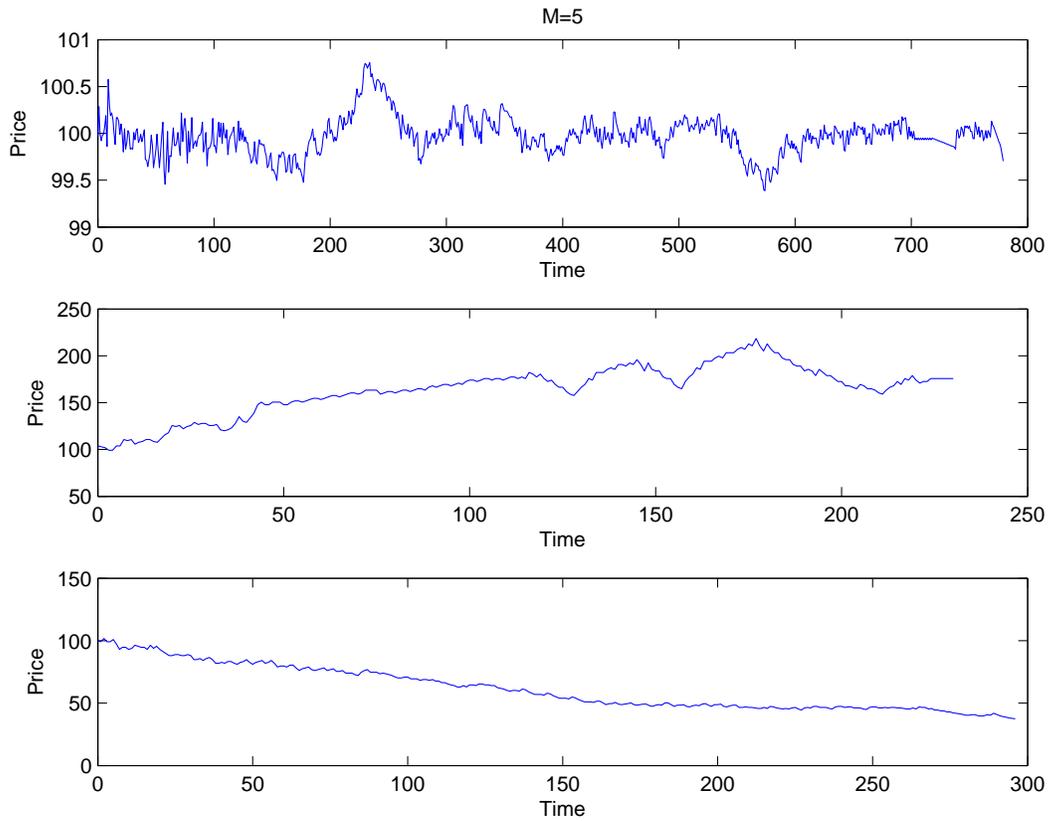}
\caption{\protect\label{Fig2}
3 different examples corresponding to speculative price behavior in a fundamental 
value/increasing/decreasing
 market.
}
\end{figure}

\begin{figure}[h]
\includegraphics[width=14cm]{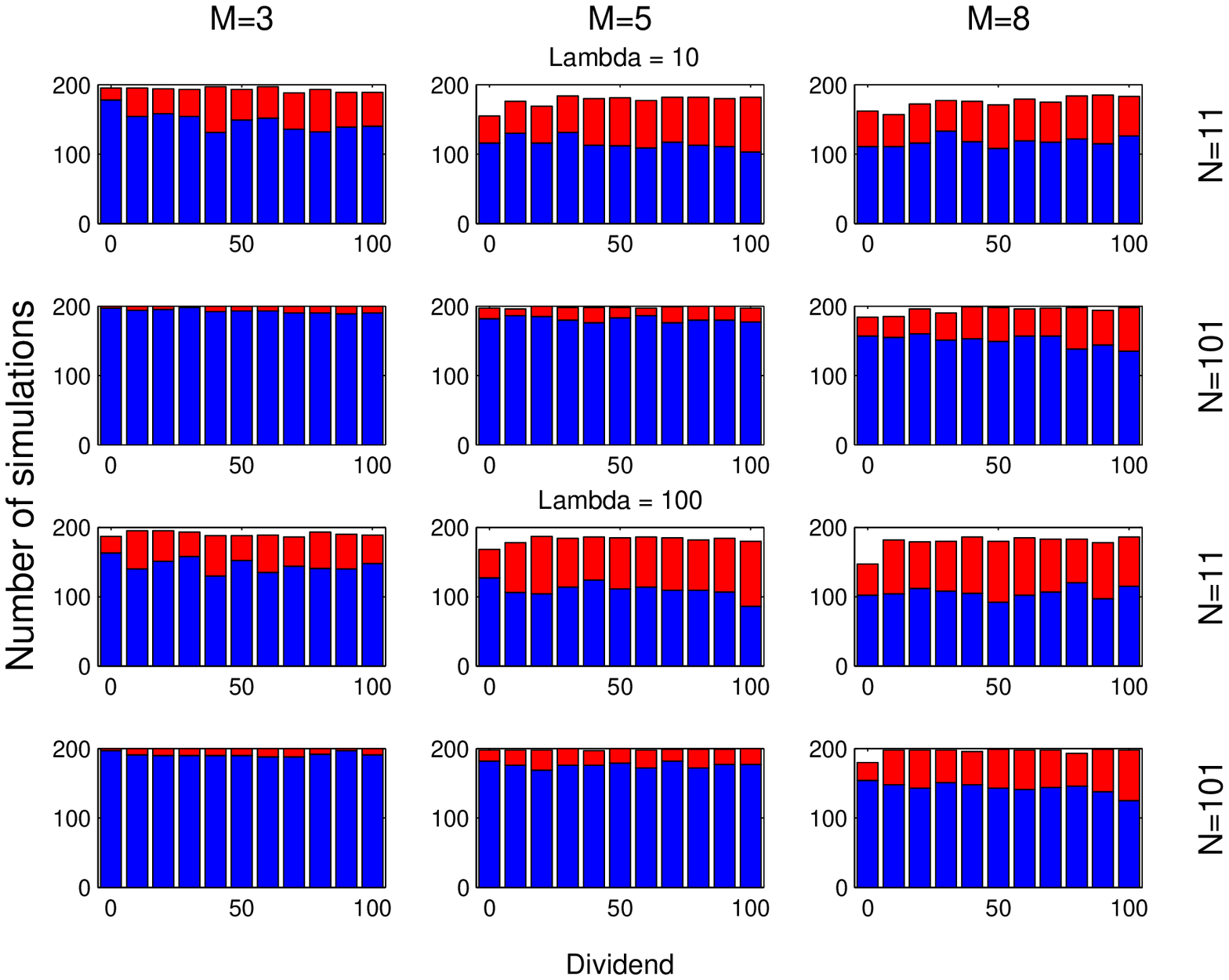}
\caption{\protect\label{Fig3}
Histograms representing respectively speculative behavior (blue) or  
fundamentalst  behavior (red) as outcomes in a setup of the \$-Game for $s=2$ with given 
parameter values of $(N,m,\lambda,d)$.
}
\end{figure}

\begin{figure}[h]
\includegraphics[width=14cm]{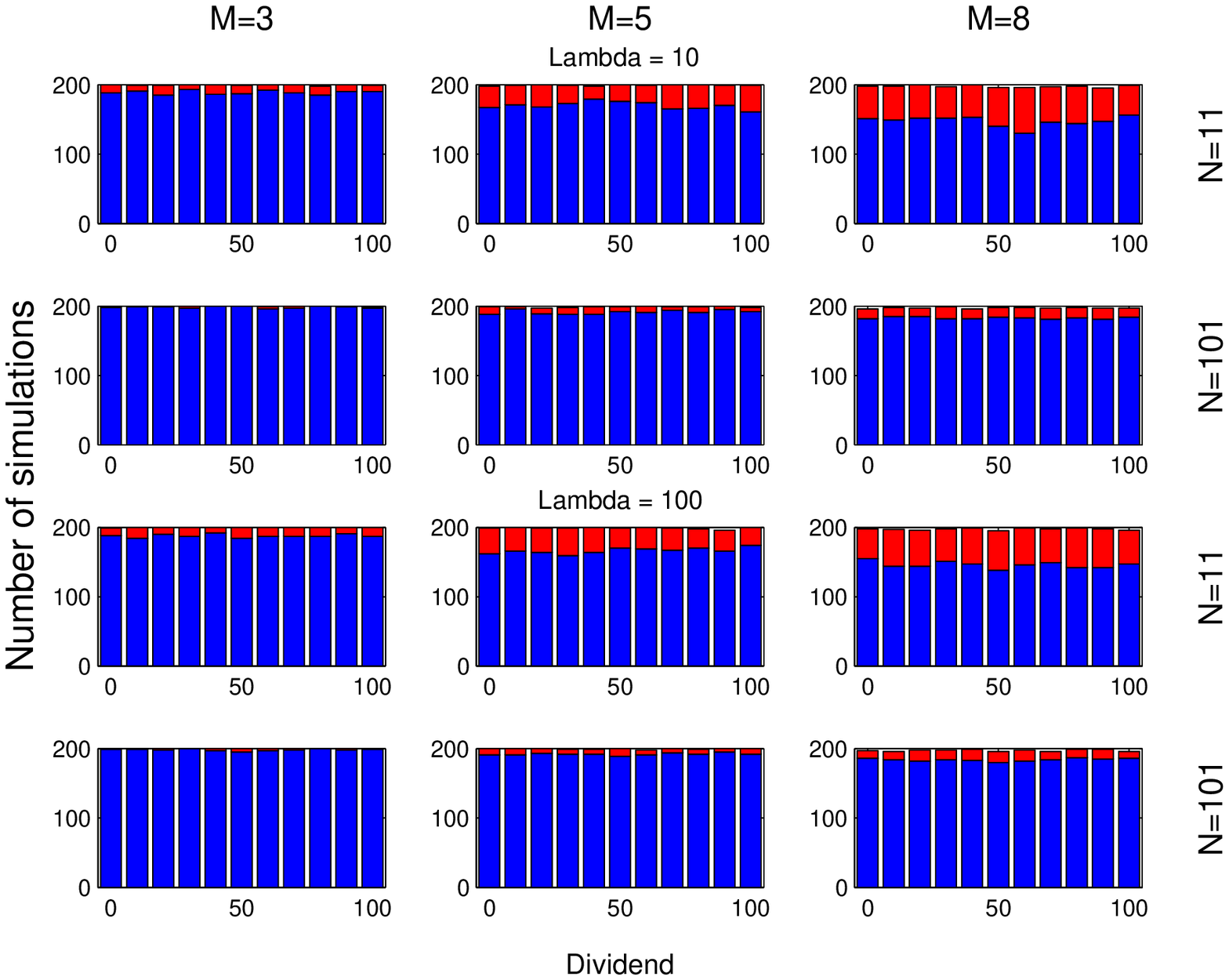}
\caption{\protect\label{Fig4}
Histograms representing respectively speculative behavior (blue) or  
fundamentalst  behavior (red) as outcomes in a setup of the \$-Game for $s=18$ with given 
parameter values of $(N,m,\lambda,d)$.
}
\end{figure}

\end{document}